\documentclass[fleqn,usenatbib]{mnras}

\usepackage{newtxtext,newtxmath}

\usepackage[T1]{fontenc}

\DeclareRobustCommand{\VAN}[3]{#2}
\let\VANthebibliography\thebibliography
\def\thebibliography{\DeclareRobustCommand{\VAN}[3]{##3}\VANthebibliography}

\usepackage{graphicx}	
\usepackage{amsmath}	
\usepackage{svg}

\title[QPO detection in 4FGL 2022.7+4216]{Detection of $\sim \text{100 days}$ periodicity in the gamma-ray light curve of the BL Lac 4FGL 2022.7+4216}

\author[Banerjee et al.]{
Anuvab Banerjee,$^{1}$
Ajay Sharma,$^{1}$\thanks{E-mail: ajjjkhoj@gmail.com }
Avijit Mandal,$^{1}$
Avik Kumar Das,$^{2}$
Gopal Bhatta,$^{3}$
and Debanjan Bose$^{4}$
\\
$^{1}$S. N. Bose National Centre for Basic Sciences, Block JD, Sector III, Salt Lake, Kolkata 700 106, India\\
$^{2}$Department of Physical Sciences, Indian Institute of Science Education and Research Mohali,
Knowledge City, Sector 81, SAS Nagar, Punjab 140306, India\\
$^{3}$Institute of Nuclear Physics, Polish Academy of Sciences, PL-31342 Krak\'{o}w, Poland\\
$^{4}$School of Astrophysics, Presidency University, 86/1 College Street, Kolkata 700073, West Bengal, India  
}

\date{Accepted XXX. Received YYY; in original form ZZZ}

\pubyear{2023}

\begin{document}
\label{firstpage}
\pagerange{\pageref{firstpage}--\pageref{lastpage}}
\maketitle

\begin{abstract}
Study of quasi-periodic oscillations (QPO) in blazars is one of the crucial methods for gaining insights into the workings of the central engines of active galactic nuclei. QPOs with various characteristic time scales have been observed in the multi-wavelength emission of blazars, ranging from the radio to gamma-ray frequency bands. In this study, we carry out a comprehensive variability analysis of the BL Lac object 4FGL 2022.7+4216 detected by the \textit{Fermi-}LAT, over a period of more than three years, from April 27, 2019 to August 09, 2022. By utilizing multiple widely-used methods of time-series analyses, we detect the presence of quasi-periodic fluctuations with a period of $\sim$100 days  with a confidence level exceeding $4\sigma$. This is the first time such a variability feature pertaining to this source is being reported. We propose that the observed QPO may be related to the precession of the blazar jet with a high Lorentz factor or to the motion of a plasma blob through the helical structure of the jet. However, for a decisive conclusion on the physical origin of such fluctuation, further multi-wavelength complementary observations, especially Very Long Baseline Interferometric observations, would be required. 
\end{abstract}




\section{Introduction}

Blazars are a subclass of active galactic nuclei (AGN) with their relativistic jets pointed toward our line of sight. Blazar continuum emission exhibits rapid flux variability over a wide range of time scales, such that multifrequency variability studies offer significant insights into the location, size, and underlying physical processes of the emission regions. Moreover, as the high energy $\gamma$-ray emission of blazars originate from their jets, the exploration of $\gamma$-ray variability provides clues regarding the jet dynamics of such systems. \par 
Even though the blazar variability is often of stochastic nature such that the power spectral density can be fairly represented by a power-law over a wide range of temporal frequencies\citep[see][and the references therein]{2020ApJ...891..120B,Sobolewska_2014}, there have been claims of the detection of quasi-periodic oscillations (QPO) in different wavebands with a time scale ranging from a few days to years \citep[e.g.][]{rani2009nearly,bhatta2019blazar,gupta2019detection,2021ApJ...923....7B}. In particular, since the $\gamma$-rays in blazars are widely believed to be originated from the highly collimated relativistic jets, QPO detection in $\gamma$-ray is crucial to infer the jet dynamics and particle acceleration mechanisms. It was, however, pointed out that many of such detection last for only 2-4 cycles and the significance of those detection is likely to be overestimated \citep{gupta2014JApA...35..307G}.\par
Owing to the continuous monitoring of \textit{Fermi-}LAT to explore the $\gamma$-ray sky, a few significant detection of $\gamma$-ray QPOs have been, however, reported; namely, 34.5-day QPO in PKS 2247-131 \citep{zhou2018NatCo...9.4599Z}, $\sim$47-day QPO in 3C454.3 \citep{sarkar2021MNRAS.501...50S}, and $\sim$7.6-day QPO in CTA 102 \citep{sarkar2020A&A...642A.129S}. Using a systematic search for periodicities using a comprehensive likelihood estimation on a large sample of $\gamma$-ray detected blazars, \citet{penil2020ApJ...896..134P} found only 11 sources showing strong signatures of periodicity of $>4\sigma$ significance. \par

The BL Lacerate object 4FGL J2202.7+4216 at redshift $z = 0.069$ was detected in a high flux state on May 1, 2019, with a daily averaged flux reaching $(1.5 \pm 0.2) \times 10^{-6} ~\text{photons cm}^{-2}\text{s}^{-1}$ by \textit{Fermi-}LAT \citep{2019ATelGarrappa}. Subsequently, on August 19, 2020, \textit{Fermi-}LAT has captured the source with daily averaged flux $(2.8 \pm 0.2) \times 10^{-6} ~\text{photons cm}^{-2}\text{s}^{-1}$ \citep{2020ATelOjha}. The $\gamma$-ray flaring activity has coincided with the optical brightening where R magnitude $< 12$ has been registered by several optical monitoring campaigns \citep{2020ATelOpticalGrishina,2020ATelATOM,2020ATelSteineke}. Strong intra-night optical variability has been detected by Automatic Telescope for Optical Monitoring (ATOM) monitoring program \citep{2020ATelATOM}. The detection of a 356 GeV photon with a strong possibility to be associated with the source has also been reported \citep{2019ATelGarrappa}. Therefore, it can be regarded as a plausible candidate for the very high energy (VHE) emitting source. \par 

In this work, we present a study the Fermi/LAT gamma-ray light curve of the BL Lac object 4FGL 2022.7+4216 detected by the \textit{Fermi-}LAT spanning more than three years. By employing multiple methods of time-series analyses, we report  presence of QPO with a characteristic timescale of $\sim$100 days. In Section 2, we provide an outline of the data acquisition and processing method used for the Fermi/LAT telescope, including relevant details. In Section 3,  analysis and the results of time series methods, that is, LSP and WWZ and auto-regressive method, are presented. In Section 4, the result and its implications are discussed in the light of standard model of AGN and conclusions are summarized.

\section{OBSERVATIONS AND DATA REDUCTION}

\begin{figure*}
    \centering
   \includegraphics[width=14cm,height=6cm]{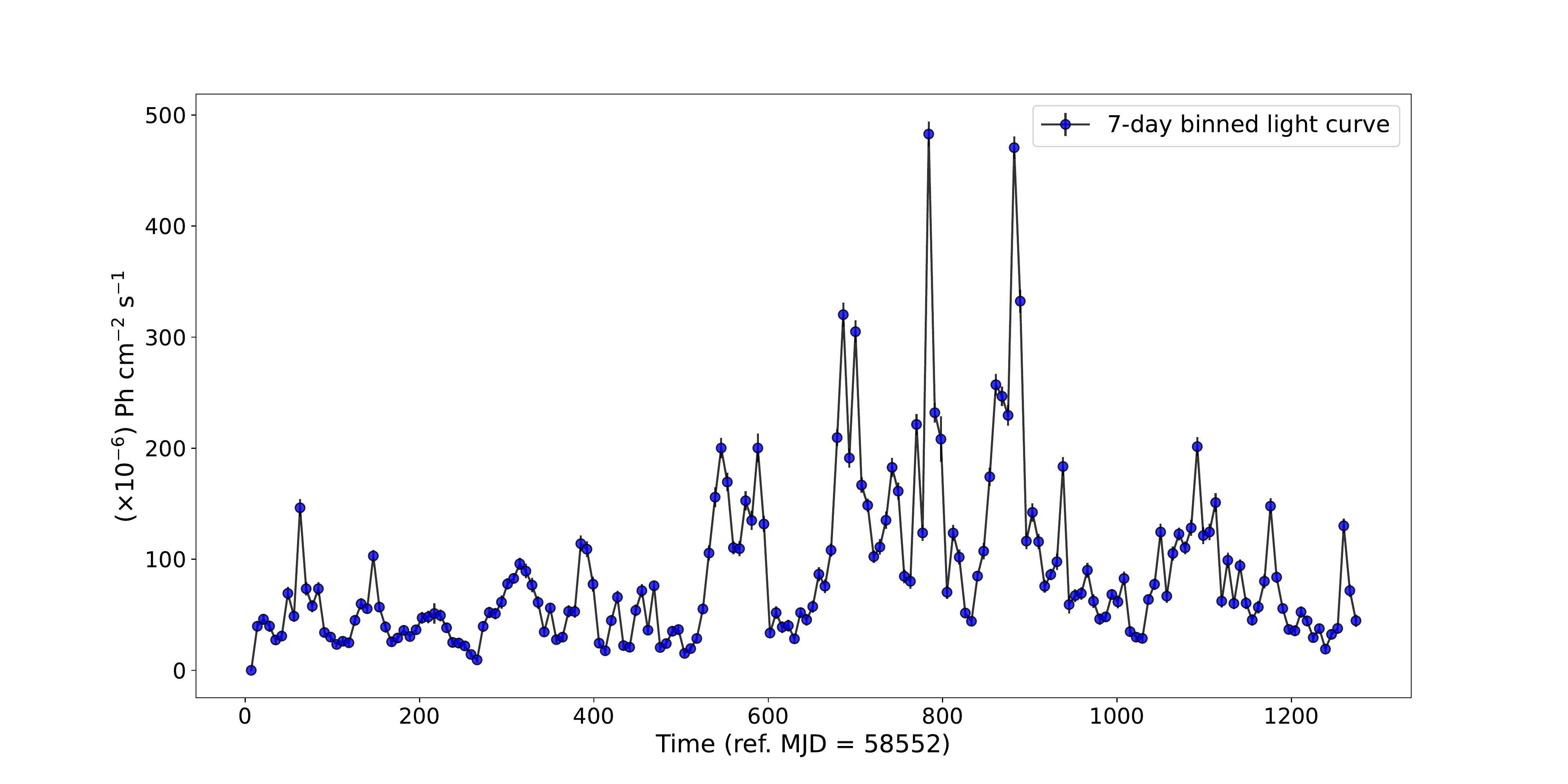}
    \caption{The Fermi-LAT light curve of the blazar 4FGL2022.7+4216 with 7 day binning (TS > 25) in MJD 58552-59847. }
    \label{fig-1}
\end{figure*}

A study of the source $4FGL 2022.7+4216$ has been done using the Fermi-LAT data. Fermi is a space-based gamma-ray observatory onboard two instruments (One is the Large Area Telescope (LAT) and another is Gamma-ray Burst Monitor (GBM)). The LAT has a large effective area of > 8000 $cm^2$ at $\sim 1GeV$. It is a pair conversion detector, that mainly operates in the energy range 20 MeV - 300 GeV and has a FOV of 2.4 sr \citet[]{atwood2009large}, covering about 20$\%$ of the sky at any time, and the whole sky in every three hours.\\
In this study, the source was analyzed in the time domain MJD 58552-59847 and was selected with 15$^{\circ}$ circular region of interest(ROI) centered at RA: 330.68 and Dec: 42.2778.\\
The recommended Fermi science tools `FERMITOOLS' package was used to analyze the Fermi-LAT data of the source. The data was obtained in the energy range 20 MeV - 300GeV from the fermi PASS 8 database and it was filtered using \textsc{gtselect} FERMITOOLS tool with constraints evclass=128 and evtype=3 and to prevent the source contamination with from earth limb, we set a criterion on the zenith angle, should be less than 90$^{\circ}$. We filtered the data with the standard filter $'\text{(DATA\_QUAL > 0) \&\& (LAT\_CONFIG == 1)}'$ using the \textsc{gtmktime} tool to obtain high-quality data in the good time intervals (GTIs). The integrated livetime as a function of sky position and off-axis angle and exposure were computed using \textsc{gtltcube} and \textsc{gtexposure} tasks respectively. An unbinned likelihood analysis was performed using the \textsc{gtlike} tool \citep{cash1979parameter,mattox1996likelihood} which provided the significance of each source within the region of interest (ROI) in the form of test statistics (TS $\sim \sigma^2$) \citep{mattox1996likelihood}. The 7-day binned light curve was obtained by integrating the source fluxes for the intervals where $\text{TS} > 25$ (above $\sim 5 \sigma $ significance; Figure-\ref{fig-1}). The diffuse $\gamma$-ray emissions of the galactic and extra-galactic were modeled using two files : $\text{gll\_iem\_v07.fits}$ and $\text{iso\_P8R2\_SOURCE\_V3\_v1.txt}$ and adopted instrument response functions were $\text{P8R2\_SOURCE\_V3 }$ to obtain the spectrum of the source. The iterative fitting of the light curve was done using $\text{ENRICO}$ software \citep{sanchez2013enrico}.

\begin{figure*}
    \centering
    \includegraphics[width=0.9\linewidth]{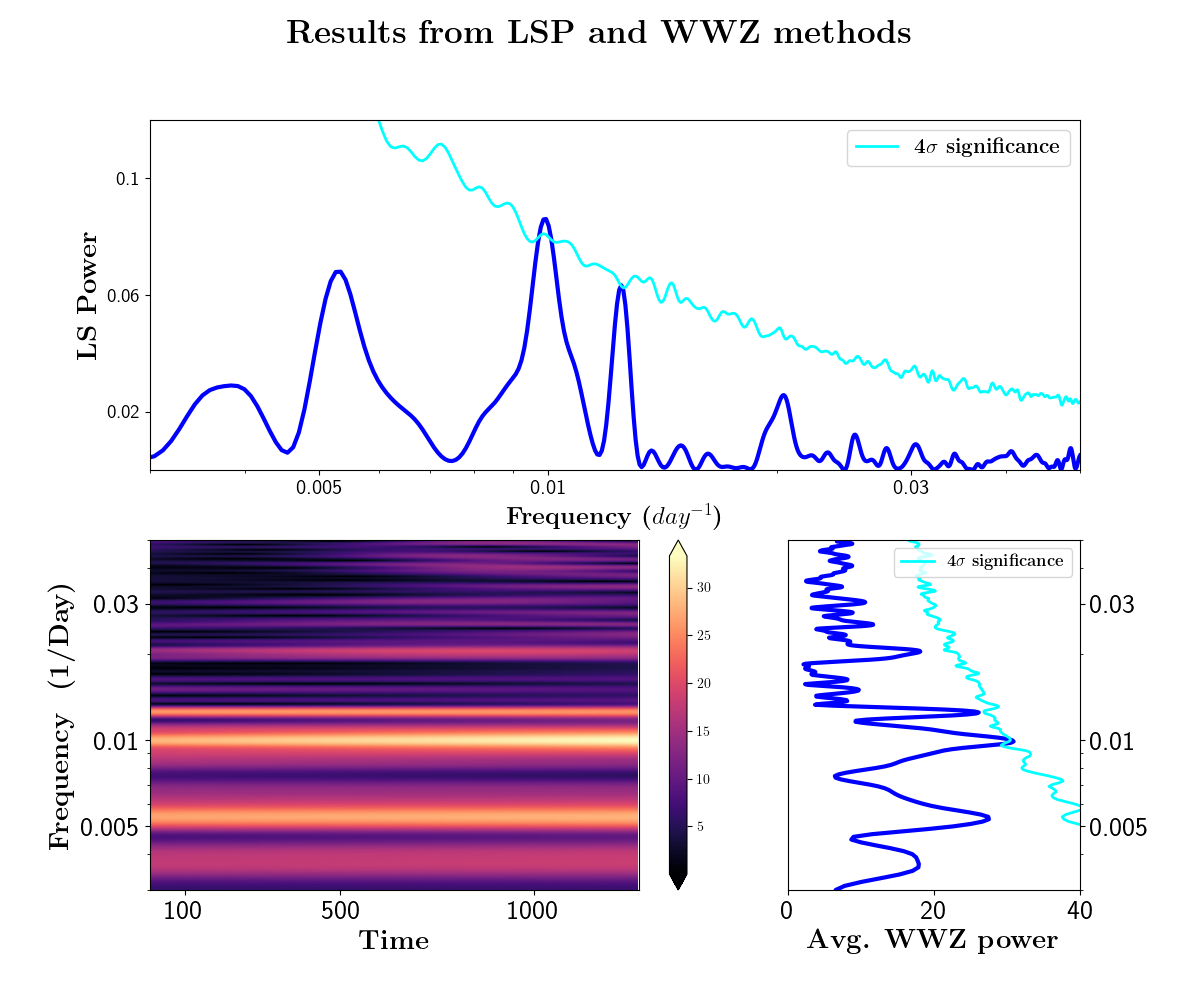}
    \caption{Top: The Lomb-Scargle periodogram corresponding to the lightcurve and the 4$\sigma$ significance line. The period around $0.01~\text{days}^{-1}$ happens to be $>4\sigma$ significant. Bottom left: WWZ map of the source 4FGL2022.7+4216 light curve in MJD 58552-59847, colors in color bar are indicating the wavelet power levels. The decipherable power concentration around $0.01~\text{days}^{-1}$ indicates the plausibility of the presence of a periodicity. The time stamps are relative to MJD = 58552. Lower right panel: Blue solid line is representing the average wavelet power across time and the cyan solid line marks the $4\sigma$ significance level of average WWZ power. The statistical significance of $0.01~\text{days}^{-1}$ is corroborated by this method as well.}
     \label{fig-2}
\end{figure*}



\section{DATA ANALYSIS \& RESULTS}

We adopted several quantitative tests to search for QPO in the gamma-ray light curve of the source, namely the `Weighted Wavelet Z-transform (WWZ)' method, `Lomb-Scargle Periodogram (LSP)' method, and `REDFIT' method. Below we describe the results obtained by the application of these methods. 

\subsection{Lomb-Scargle Peridogram}

The LSP, introduced by Lomb \citep{lomb1976least} and extended later by Scargle \citep{scargle1982studies}, is actually a variant of the traditional discrete Fourier transform (DFT). However, it has a distinct advantage over DFT, which is that for uneven sampling, it reduces the effect of sampling irregularities by iteratively fitting sinusoidal waves to the data. We compute the LSP using the \textsc{astropy LombScargle} class by considering the minimum and maximum temporal frequencies to be $1/T$ and $1/2\Delta{t}$ respectively, where $T$ is the total observation period and $\Delta{t}$ is the time-binning. The resulting LSP shows a prominent peak at $0.01~\text{days}^{-1}$. \par 
However, it is quite well known that variability observed in blazar lightcurves are also associated with underlying red noise, which can lead to apparent periodic behavior for a few cycles within the low-frequency regime \citep{vaughan2005simple}. We, therefore, attempt to determine the statistical significance of the periodic feature using the method proposed by \citet{emmanoulopoulos2013generating}. We approximate the observed power spectra by a power-law model and simulate 1000 light curves with the best-fit power spectral slope and flux distribution as that of the original light curve. The periodicity feature at $0.01~\text{days}^{-1}$ is found to be $>4\sigma$ significant (Top panel of Figure \ref{fig-2}).  \par 
Furthermore, it is conventional to test the presence of periodicity using Generalized
Lomb-Scargle Periodoram (GLSP)\footnote{\url{https://pyastronomy.readthedocs.io/en/latest/pyTimingDoc/pyPeriodDoc/gls.html}} as well which accounts for the measurement errors in the analysis. We obtain the periodicity with the same $\sim$100 day period using GLSP as well, which further strengthens our claim. \par 

Statistically speaking, however, since we do not have a priori expectations of having a particular periodicity, it is more rigorous to estimate the `global significance', which is the fraction of surrogate light curves showing a LSP peak larger than the observed confidence level at any frequency value. In this way, it is ensured that the peaked component is searched over a larger frequency interval, and consequently, we are sampling from a larger population of `false positive' signals. If we do not restrict the peak to reside within a particular frequency range, then we find that the peaked component at 100 days is associated with $>99\%$ global confidence. 

\subsection{Weighted Wavelet Z-Transform}

The standard LSP method has the limitation that it attempts to fit the sinusoidal profile across the entire domain of observation and does not account for the fact that the features coming from the real astrophysical observation could be time-dependent, i.e. the amplitude and frequency could evolve over time. Therefore, in order to characterize the periodicity features and their evolution, the wavelet transform method turns out to be a more suitable tool, which convolves the light curve with the time and frequency-dependent kernel and attempts to localize the periodicity feature in time and frequency. For the purpose of our analysis, we use the Morlet kernel \cite{grossmann1984decomposition} with the functional form:

\begin{equation}
    f[\omega (t - \tau)] = \exp[I \omega (t - \tau) - c \omega^2 (t - \tau)^2]
\end{equation}

and the WWZ map is given by,
\begin{equation}
    W[\omega, \tau: x(t)] = \omega^{1/2} \int x(t)f^* [\omega(t - \tau)] dt
\end{equation}
We use publicly available software\footnote{\url{https://github.com/eaydin/WWZ}} to estimate the WWZ power as a function of frequency and evolution of time. In the time-frequency plane, the color-scaled WWZ map demonstrates a decipherable concentration of power around $0.01~\text{days}^{-1}$, which is apparent in the average WWZ power as well  (Bottom left panel of Figure \ref{fig-2}). We estimate the significance of this peak WWZ power in the same way by simulating 1000 light curves according to the best-fit PSD model and resampling the simulated light curves according to the source light curves. The peak significance was found to be $>4\sigma$ here as well (Bottom right panel of Figure \ref{fig-2}).

\begin{figure}
    \centering
    \includegraphics[scale=0.45]{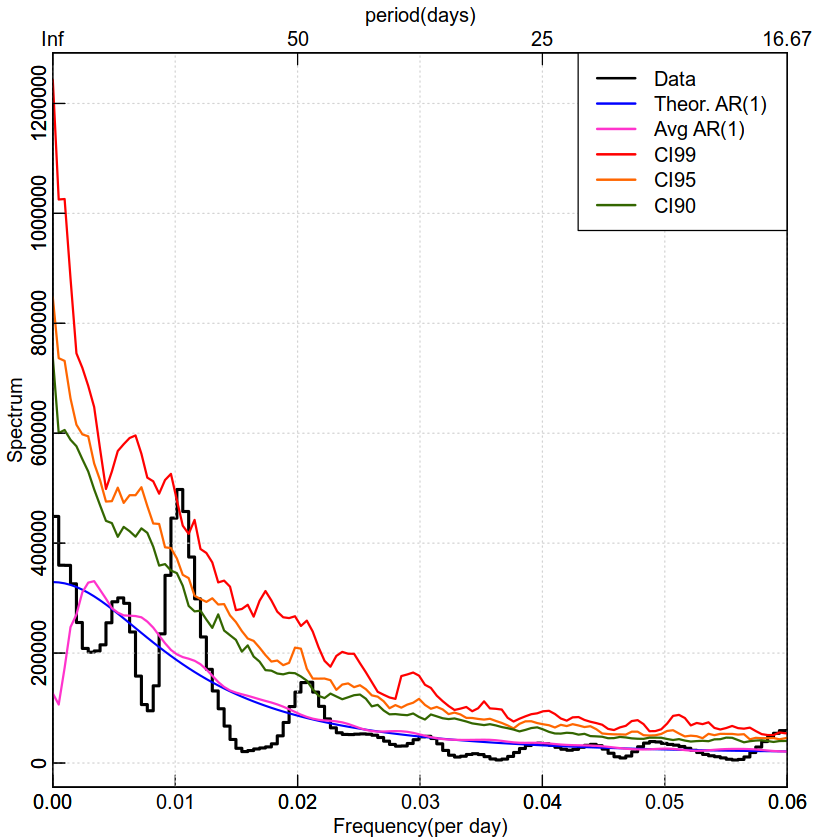}
    \caption{Power spectrum and the significance levels using REDFIT method. The blue and magenta curves denote the theoretical AR(1) spectrum and the mean simulated AR(1) spectrum respectively. The green, orange and red curves represent 90, 95, and 99\% significance levels respectively.}
    \label{fig-redfit}
\end{figure}

\subsection{REDFIT}

In the REDFIT method, the unevenly spaced time series data are fitted with a first-order autoregressive process (AR1), avoiding interpolation and its inevitable bias in the time domain \citep{schulz2002redfit}. This method is also used to test the significance of the flux peaks in time series against the background of red-noise in the first-order autoregressive process. The usage of AR1 process is justified by the autoregressive nature of the emission flux of blazars \citep{schulz2002redfit, kushwaha2020possible}, where the current emission flux is dependent on its previous flux state. We use a REDFIT program to  estimate the spectrum using LSP (Lomb-scargle periodogram) and WOSA (Welch-overlapped-segment-averaging) procedures with the number of WOSA segments ($n_{50}=1$); a Welch window was chosen to reduce spectral leakage. The bias-corrected power spectrum alongside the theoretical and simulation-generated AR(1) processes are provided in Figure-\ref{fig-redfit}. It shows a prominent peak of about 100 days with a confidence level of 99$\%$ (red curve), which is the maximum significance provided by the REDFIT program.

\section{DISCUSSION AND CONCLUSIONS}
In this work, we report the detection of 100 days periodicity pertaining to the $\gamma$-ray detection BL Lac object 4FGL 2022.7+4216 using three different methods of time series analysis. Below we illustrate a few physical scenarios which can give rise to the periodic behaviour of $\gamma$-ray lightcurve and thereby infer the plausible mechanisms operative in our context on the basis of time scale.
\begin{itemize}
    \item In the case of binary supermassive black hole (SMBH) systems, if the secondary black hole pierces the accretion disk of the primary black hole, QPO may be observed as a consequence of the impact flashes \citep{valtonen2008massive}. Such a QPO of a period of $\sim$12 years has been reported earlier in the context of OJ 287 \citep{valtonen2008massive}. However, the time scale corresponding to this process is $\sim$ years, and, therefore, the present detection of 100-day periodicity is difficult to explain under the purview of this scenario. 
    \item Since blazar emission is dominated by jets, it is quite likely that the QPO signature will have some bearing on the jet emission features. The precession of the blazar jets could be induced by the interaction of the secondary source and could manifest itself as quasi-periodicity owing to the varying Lorentz factor \citep{begelman1980massive}. Furthermore, the orientation of the jet could also be influenced by the Lense-Thirring precession of the inner edge of the disk. However, such mechanisms typically result in periodicity $\sim$ 1-2 years time-scale \citep{2007ralc.conf..276R} and fall outside of the ballpark of the present detection. However, for blazars with jets closely aligned with our line of sight, the detected periodicity could be significantly shorter because of the Doppler boosting effect \citep{rieger2004geometrical}. Such a mechanism has been inferred to explain the 34.5-day QPO in the case of BL Lac PKS 2247–131 \citep{zhou201834}. In our case, therefore, such a mechanism could be operative if the jet is associated with large Lorentz factors. 
    \item The variable Doppler factor arising out of the movement of the plasma blob along the internal helical structure of the jet could be another plausible physical mechanism of jet-driven quasi-periodicity \citep{camenzind1992lighthouse}. Such helical structures could come from the interaction of the jet with the surrounding medium \citep{godfrey2012periodic} or the hydrodynamic instability effects \citep{hardee1999dynamics}. Depending upon the parameters like the pitch angle, the viewing angle, and the Doppler boosting factor, the variability time-scale can range from $\sim$ a few days to $\sim$ months \citep{rani2009nearly}. The observed 100-day periodicity could be a result of such structural effects of the jet. 
\end{itemize}

Given the time scale of the QPO, we infer it is most likely originating from a precessing jet with a high Lorentz factor or because of the motion of the plasma blob along a curved jet. In the case of the one-zone leptonic model where the plasma blob moves along the postulated helical trajectory, the time dependence of the viewing angle will cause varying Doppler factor and consequent intensity variation even without the intrinsic changes in jet emission. The time dependence of the viewing reads as
\begin{equation}
    \cos{\theta_{\text{obs}}(t)} = \sin{\phi}\sin{\psi}\cos{2\pi{t}/P_{\text{obs}}} + \cos{\phi}\cos{\psi}
\end{equation}
where $\psi$ is the jet angle relative to our line of sight and $\phi$ is the pitch angle of the blob.  $P_{\text{obs}}$ is the observed periodicity \citep{sobacchi2016model}. The Doppler factor varies with time as $\delta(t) = 1/(\Gamma(1-\beta\cos{\theta(t)}))$, where $\Gamma = 1/\sqrt{1-\beta^2}$ is the bulk Lorentz factor of the jet motion where $\beta = \frac{v_{\text{jet}}}{t}$. The periodicity in the rest frame of the blob is estimated by
\begin{equation}
    P_{\text{rf}} = \frac{P_{\text{obs}}}{1 - \beta\cos{\psi}\cos{\phi}}.
\end{equation}

 For typical values of $\phi = 2^{\circ}$, $\psi = 5^{\circ}$ and $\Gamma = 8.5$, the rest frame periodicity becomes $\sim$24 years for $P_{\text{obs}} \sim 100$ days as we have obtained in our case. During one period, the blob traverses a distance $D = c\beta P_{\text{rf}}\cos{\phi} \sim 7$ parsec. Since the prominent QPO signature of $\sim$100 days is observed throughout the entire domain of $\sim$1200 days in the WWZ map, we expect roughly $\sim$12 cycles of oscillations during this period. Therefore, the projected distance during these 12 cycles would be estimated as $D_p \sim 12D\sin{\psi} = 7.2$ pc. Such parsec scale helical jets have been found in the case of several other blazar sources as well \citep{vicente1996A&A...312..727V,Tateyama2002ApJ...573..496T}.
 
 \citet{roy2022transient} considered that the inclination angle of the jet axis relative to the line of sight ($\psi$) could be time-dependent, which can explain the time variation of the amplitude of oscillation. Therefore, the amplitude variation would then be a direct offshoot of the geometric bending structure of the jet. However, a more comprehensive investigation, including detailed very long baseline interferometry (VLBI) monitoring needs to be undertaken to confirm the presence of such curvature within a length scale of $7$pc.


\section*{Acknowledgements}

This study made use of the \textit{Fermi}–LAT data,
obtained from the Fermi Science Support Center (FSSC), distributed by
NASA's Goddard Space Flight Center (GSFC).
D. Bose acknowledges the support of Ramanujan Fellowship-SB/S2/RJN-038/2017.
A. Sharma and A. Mandal are grateful to S. N. Bose National Centre for Basic Sciences under the Department of Science and
Technology (DST), Government of India, for providing the necessary
support to conduct this research.

\section*{Data Availability}
The work uses publicly available data from \textit{Fermi-}LAT.



\bibliographystyle{mnras}
\bibliography{example} 







\bsp	
\label{lastpage}
\end{document}